\documentstyle[prd,aps]{revtex}
\begin{document}
\draft
\twocolumn[\hsize\textwidth\columnwidth\hsize\csname
@twocolumnfalse\endcsname
\preprint{
UAB-FT-406,
UMD-PP-97-62,
Fermilab-PUB-96/433-A, 
Submitted to Phys. Rev. Lett.
{}~hep-ph yymmnn}
\title{New Astrophysical Constraints on the Mass of the 
Superlight Gravitino}
\author{J. A. Grifols$^{1}$, R. N. Mohapatra$^{2}$ and
A. Riotto$^{3}$}
\address{$^{(1)}${\it  Grup de F\'{\i}sica Te\`orica and IFAE,
Universitat Aut\`onoma de Barcelona,
08193 Bellaterra,  Spain.}}
\address{$^{(2)}${ Department of Physics, University of Maryland, 
College Park, Md-20742, USA.}}
\address{$^{(3)}${Fermilab
National Accelerator Laboratory, Batavia, Illinois~~60510}}
\date{Dec. 4, 1996}
\maketitle
\begin{abstract}

In some supergravity models, the superlight gravitino is accompanied by
a light weakly coupled scalar ($S$) and pseudoscalar particle ($P$). The
couplings of these particles to matter (e.g. electrons and photons)
is inversely proportional to the product $(m_{\tilde{g}}M_{P\ell})$
where $m_{\tilde{g}}$ and $M_{P\ell}$ are respectively the  
gravitino mass and the Planck mass. As a result, their emission from 
supernovae and stars via the reaction $\gamma + e^- \to S/P + e^-$
for certain ranges of the gravitino mass can become the dominant energy
loss mechanism in contradiction with observations thereby ruling out
those mass values for the gravitino. For 10 MeV$\geq m_{S/P}\geq $ keV, the 
SN1987A observations can be used to exclude 
the gravitino masses in the range, $ (10^{-1.5} \leq
m_{\tilde{g}} \leq 30)$ eV whereas if
$m_{S/P}\leq$ keV, constraints of stellar energy loss can exclude the
range $(3\times 10^{-6} \leq m_{\tilde{g}}\leq 50)$ eV for the photino mass
equal to 100 GeV. We also find that if $m_{S/P}\leq $ MeV, 
present understanding of Big Bang Nucleosynthesis 
imply that $m_{\tilde{g}}\geq $ eV. These are the most severe bounds
to date on $m_{\tilde{g}}$ in this class of models.                    

\end{abstract}
\pacs{  UAB-FT-406 \hskip 1cm UMD-PP-97-62 \hskip 1cm 
FERMILAB--Pub--96/433-A \hskip 1 cm hep-ph/9612253}

 \vskip2pc]

In generic supergravity models the gravitino ($\tilde{g}$) mass 
is given by the formula 
$m_{\tilde{g}}\sim\frac{ \Lambda^p_{\rm SUSY}}{ M^{p-1}_{P\ell}}$
where $\Lambda_{\rm SUSY}$ is the scale of supersymmetry breaking and
$M_{P\ell}$ is the Planck mass that characterises the gravitational
interactions.
Any information on the gravitino mass therefore translates into knowledge
of one of the most fundamental parameters of particle physics, the scale
at which supersymmetry breaks. In this letter, we consider  
a class of supergravity models, which have a superlight gravitino, $\tilde{g}$
and discuss new astrophysical constraints on the allowed mass range for the
gravitino, when it is accompanied by a superlight scalar ($S$) 
and pseudo-scalar ($P$) particle. 

Typical example of models where a superlight gravitino
may arise are the so-called GMSB models, where 
gauge interactions mediate the breakdown of 
supersymmetry \cite{dine} or the class of models where
an anomalous $U(1)$ gauge symmetry induces SUSY breaking, (provided
one chooses the SUSY breaking scale to be in the TeV range 
rather than $10^{11}$ GeV considered in several examples \cite{dvali}).
In these models, the above formula for $m_{\tilde{g}}$ comes with a value
for $p=2$ and the low value for the scale $\Lambda_{{\rm SUSY}}$ 
then tells us that the mass of the gravitino
in this class of models is anywhere between $10^{-6}$ eV to a keV.   
There is also a class of no-scale models \cite{ellis} where gravitinos can be
superlight. In all these models, gravitino is
supposed to be the lightest supersymmetric particle (LSP). Not all these
models will have superlight $S/P$ particles accompanying the gravitino;
however, in a subclass of them\cite{roy}, $\tilde{g}$ as well as 
$S/P$ will remain superlight. These particles as well as the gravitino 
are coupled to visible matter via couplings which are inversely 
proportional to the product $(m_{\tilde{g}} M_{P\ell})$\cite{roy2}.
Therefore for superlight gravitinos, the Planck mass suppression in the
couplings can be overcome by the small value of $m_{\tilde{g}}$
leading to an enhancement of their production in both colliders \cite{nandi}
as well as astrophysical and cosmological setting such as stars \cite{nowa},
supernovae \cite{grifols} and the early universe \cite{gherghetta}.
They also lead to enhanced contributions to low energy parameters such
as the $g-2$ of the muon \cite{mendez}. Presently available information
in particle physics as well as astrophysics can therefore be used to
constrain the mass of the gravitino, which in turn can be used to
gain information on the scale of supersymmetry breaking.

In a recent paper\cite{grifols}, 
we showed that if the gravitino is the only superlight
particle in the theory, (or $m_{S/P}\geq 100$ MeV) 
observed supernova neutrino luminosity by the IMB and Kamiokande
groups \cite{bionta} and its understanding in terms of the standard model
of the supernova \cite{burrows} 
allows us to exclude the range of its mass between $10^{-8}$ eV to
$10^{-6}$ eV. The main production channel for gravitinos responsible
for draining energy away from the neutrinos (and hence leading to this
bound) turns out 
to be  $\gamma\gamma$ collision. On the other hand if the gravitino
is accompanied by superlight $S/P$ particles, then $S/P$ can be produced
via the reaction $\gamma + e^-\to S/P + e^-$ opening up a new channel
for energy loss. It is the purpose of this
paper to report on our study of how the presence of this reaction channel
effects the photon emission from supernovae, stars and the considerations of
big bang nucleosynthesis (BBN). 

In the case of the supernova, we require first
that the emission of the $S/P$ particles do not drain more than $10^{52}$
ergs/sec of the energy, which enables us to put a lower bound 
$m_{\tilde{g}}\geq 30$ eV if $m_{S/P}\leq 10 MeV$. 
It however turns out that as $m_{\tilde{g}}$
goes below $0.3$ eV, its mean free path becomes less than the radius
of the supernova core and the usual energy loss discussion does not apply. 
An extension of this excluded range can then be obtained from
energy loss from Red Giants and the Sun considered in Ref. \cite{nowa} 
and reconsidered here to give  
$(10^{-3.5}\leq m_{\tilde{g}}\leq 50)$ eV. We show that further restrictions
on $m_{\tilde{g}}$ can be obtained by using the recent result of Kolb,
Mohapatra and Teplitz \cite{kolb} which says that if the mean free path
of the photon is more than ten times that of the $S/P$ particles, then the
photon luminosity of stars will be severely depleted. Requiring that
the photon luminosity depletion does not conflict with observations, 
we find the new excluded range on $m_{\tilde{g}}$ to stretch further
down to $3\times 10^{-6}$ eV.
Combining all these results then enables us to derive the excluded
range for $m_{\tilde{g}}$ for this special class of models with
superlight $S/P$ to be: $(3\times 10^{-6}\leq m_{\tilde{g}}\leq 50)$ eV 
provided $m_{S/P}\leq 1~keV$.
This is the most severe lower bound on $m_{\tilde{g}}$ to date.

We also study the effect of $S/P$ production
in the BBN era of the early universe and conclude that unless the gravitino
mass is larger than an eV, the success of the big bang model in explaining
the observed primordial Helium abundance will be hard to understand. This bound
is considerably better than the one derived in \cite{gherghetta}.
We note in passing that recent interpretation of the CDF
$\gamma\gamma e^+e^-$ event in terms of a light gravitino decay \cite{kane}
of the photino seems to imply a gravitino mass less than 250 eV or so which
is allowed by our considerations.

To start our discussion let us write down the coupling of the $S/P$ particles
to the photons \cite{roy} that results from the superHiggs
mechanism  \cite{cremmer} of the supergravity theories:
\begin{eqnarray}
e^{-1}L=-\frac{\kappa}{4} \sqrt{\frac{2}{3}}\left(\frac{M_{\tilde{\gamma}}}
{m_{\tilde{g}}}\right)\left(SF^{\mu\nu}F_{\mu\nu}+P\tilde{F}^{\mu\nu}F_{\mu\nu}
\right),
\end{eqnarray}
where $M_{\tilde{\gamma}}$ is the photino mass and 
$\kappa=\sqrt{8\pi}/M_{P\ell}$. ($\tilde{F}$ is the dual of the photon field
strength.) 
We did not display the $S/P$ couplings to the electrons since they do not
carry the $(m_{\tilde{g}})^{-1}$ enhancement in their couplings and are
therefore negligible in their contribution to the process $\gamma + e^-
\to S/P + e^-$ compared to interactions shown in Eq. (1). This scattering
channel arises from the Primakoff process involving the $S/P\gamma\gamma$
vertex at one end of the Feynman diagram and the usual electromagnetic
coupling at the other.
One can then calculate the cross-section for the production $S/P$ particles
in $e^- + \gamma$ collisions to be \cite{nowa}
\begin{eqnarray}
\sigma(\gamma e^-&\to& S/P
e^-)=\frac{\kappa^2\alpha_{em}}{6}\nonumber\\
&\times&\left(\frac{M_{\tilde{\gamma}}}{m_{\tilde{g}}
}\right)^2\left(2\:{\rm ln}\left(\frac{E_{\gamma}}{m_{S/P}}\right)
+2\:{\rm ln} 2-1\right).
\end{eqnarray}
To obtain an estimate of the amount of energy lost from the supernova core 
in $S/P$ emission, we write
\begin{eqnarray}
Q_{S/P}\simeq Vn_{\gamma} n_e \sigma(\gamma e\to S/P e)E_{S/P}
\end{eqnarray}
Using $E_{S/P}\simeq 150$ MeV and $n_e\simeq 1.6\times 10^{38}$  cm$^{-3}$
and $n_{\gamma}(T)\simeq \frac{2\zeta (3)}{\pi^2} T^3$ and requiring
that $Q_{S/P}\leq 10^{52}$ ergs/sec., we find that 
$\frac{M_{\tilde{\gamma}}}{m_{\tilde{g}}}\leq 10^{9.5}$. For the photino mass
of 100 GeV, this implies $m_{\tilde{g}}\geq 30$ eV.

Since the $S/P$ can decay to two photons with a decay rate given by
$\Gamma_{S/P}\simeq \frac{\kappa^2}{96\pi}
\left(\frac{M_{\tilde{\gamma}}}{m_{\tilde{g}}}\right)^2 m^3_{S/P}$, it
is easy to check that for $M_{\tilde{\gamma}}= 100$ GeV, $S/P$ particles
heavier than 10 MeV will have decay length of about 300$R_{SN}$. We
therefore conservatively assume that the above bound holds for $m_{S/P}
\leq 10$ MeV.

In the derivation of the lower bound on the gravitino mass given above, 
we have assumed that all the $S/P$ particles 
produced escape the supernova core. To check this let us calculate the
mean free path $\lambda_{S/P}$ of the $S/P$ particles: 
$\lambda_{S/P}\sim (n_e \sigma(S/P e\to \gamma e))^{-1}$. We find
that $\lambda_{S/P}\sim 
10^{10}\left(\frac{m_{\tilde{g}}}{30~{\rm eV}}\right)^2$ cm.
Thus as long as $m_{\tilde{g}}\geq 0.3$ eV, the $\lambda_{S/P}\geq R_{SN}$
and the $S/P$ particles escape after production and our bound applies. 
Once the mass of the gravitino is below
this value ($0.3$ eV), the $S/P$ particles get trapped and form an $S/P$
sphere and the luminosity in $S/P$ depends on the radius $R_{S/P}$
of this sphere. Using the method given in \cite{barb}, we can calculate the 
$R_{S/P}$ and we find it to be $R_{S/P}\sim\left( \frac{5}{4} \kappa^2
\alpha_{em} n_c R_c\left(\frac{M_{\tilde{\gamma}}}{m_{\tilde{g}}}\right)^2
\right)^{1/2} R_c$. In this expression, $n_c$ and $R_c$ denote respectively
the core number density and the core radius.
 Here we have assumed that the density of the supernova
goes down like $\rho(R)=\rho_C(R_c/R)^3$ \cite{turner}. Now demanding that
$\frac{Q_{S/P}}{Q_{\nu}}\equiv \left(\frac{T_{S/P}}{T_{\nu}}\right)^4
\left(\frac{R_{S/P}}{R_{\nu}}\right)^2\leq 10^{-1}$, we find the allowed
range for gravitino masses to be for $m_{\tilde{g}}\leq 10^{-1.5}$ eV.
Thus SN1987A observations seem to exclude the domain of masses between
$(10^{-1.5}\leq m_{\tilde{g}}\leq 30$) eV.

To explore further restrictions on the gravitino mass, let us look at
the stellar energy loss via $S/P$ emission.
It is well-known that if light scalar/pseudo-scalar particle have
two photon couplings, then via the Primakoff process, they constribute
to energy loss in stars \cite{teplitz,raffelt}. There are two ways to get
such constraints. One is to look for the parameter domain for which
the mean free path for the $S/P$ particle is larger than the stellar radius
($\geq 10^{11}$ cm). In this case, any production of $S/P$ particle
subtracts from the photon luminosity and must therefore be a small
fraction of the observed luminosity- i.e. the rate of energy loss
$d\epsilon_{\odot}/dt\leq 17$  ergs gm$^{-1}$ sec$^{-1}$ for the Sun and
$d\epsilon_{RG}/dt\leq 10^2$ erg gm$^{-1}$ sec$^{-1}$ for the Red Giant.
This point has already been noted in \cite{nowa}. To derive the relevant
constraints for this case, we first note that
the mean free path for the $S/P$ particle is given roughly by
$\lambda_{S/P}= \frac{1}{\sigma_{Se} n_e}\sim 10^{41}\left(
\frac{m_{\tilde{g}}}{M_{\tilde{\gamma}}}\right)^2$ cm. This implies
that mean free path exceed the typical solar radius for $m_{\tilde{g}}\geq
10^{-3.5}$ eV. For this range of masses the energy loss rate is given by
(considering scattering off electrons as well as protons):
\begin{eqnarray}
d\epsilon/dt\simeq \frac{(n_e+n_p)n_{\gamma}\sigma_{\gamma}E}{\rho}
\end{eqnarray}
where $\rho$ denotes the core density. Putting in the values for the different
parameters for a typical star, one obtains the result of Ref. \cite{nowa}
that $m_{\tilde{g}}\leq 500$ eV . In the discussion of Ref. \cite{nowa},
 the effect of the
stellar plasma has not been included. Incorporating these effects
leads to the formula (see Ref. \cite{raffelt}, Eq. (5.9)) for energy loss
per unit volume of the star via S/P particles to be:
\begin{eqnarray}
Q_{S/P}\simeq \alpha_{em}\kappa^2\left(\frac{M_{\tilde{\gamma}}}{m_{\tilde{g}}}
\right)^2 T^7 I
\end{eqnarray}
where $I$ is a function which has been calculated by Raffelt\cite{raffelt} to be
1.84 for the Sun. Requiring that $V_{\odot}Q_{S/P}\leq 10^{33}$ ergs/sec.
($V_{\odot}$ being the volume of the Sun), we get $m_{\tilde{g}}\geq 50$ eV
which is a factor of 10 weaker than the bound of Ref. \cite{nowa}.
Thus the excluded mass range for the gravitinos that comes from this
discussion is $(10^{-3.5}\leq m_{\tilde{g}} \leq 50) $ eV.

Let us now turn to the second new result of this paper which constrains
the gravitino masses below $10^{-3.5}$ eV or so when the mean free path of
$S/P$ particles is less than the solar radius. In this case, 
we use the argument of
Ref. \cite{kolb}, which goes as follows: if in photon scattering off electrons
or protons, one produces a very weakly coupled particle such as the $S/P$
in addition to the photon a small fraction of the time, the large number
of photon electron collisions undergone by the photon as it random walks
its way out of the star causes depletion of the photon flux into a flux
of the weakly coupled particle (in this case $S/P$). This depletion can be
excessive unless the $S/P+e$ scattering
rate is close to that of the Compton scattering in which case back reaction
$S/P+e^-\to \gamma +e^-$ regenerates the lost photons. In terms of the
parameter $A$ defined as $A\equiv \frac{\sigma_{\gamma+e\to
S/P+e}}{\sigma_{\gamma+e}}$, the result of Ref. \cite{kolb} is that
$A\geq 0.1$. For a star, we find
\begin{eqnarray}
A\simeq\frac{\kappa^2\alpha^{-1}_{em}m^2_e}{6\pi}
\left(\frac{M_{\tilde{\gamma}}}{m_{\tilde{g}}}\right)^2\times 2{\rm ln}(E/m_{S/P}))
\end{eqnarray}
The condition $A\geq 0.1$ then translates to $m_{\tilde{g}}\leq 10^{-9}$ eV.
However, for $m_{\tilde{g}}\leq 3\times 10^{-6}$ eV, the $S/P$ will decay inside
the star to photons and will regenerate the lost photons. So the real upper 
limit from the above argument is $10^{-6}$ eV. Note that there are lower
limits $m_{\tilde{g}}\geq 10^{-6}$ eV to $10^{-4}$ eV from collider data 
\cite{nandi} and $2\times 10^{-6}$ eV 
from present $g-2$ measurements\cite{mendez} as well 
as from the supernova\cite{grifols}.
Combining these results,  we get the results announced in
the beginning that for the class of models with superlight $S/P$ particles
accompanying the superlight gravitino, 
any value of mass for the superlight gravitino below
50 eV appears to be ruled out for $M_{\tilde{\gamma}}\simeq 100$ GeV
provided the mass of the $S/P$ particles are below one keV.
This is the most stringent lower bound on the gravitino mass to date
in this special class of models. 

We also note that the existence of the scattering mode $\gamma + e^-\to
S/P + e^-$ effects the Helium synthesis in the early universe unless the
gravitino mass is in the eV range or more. To see this let us compare the
production rate of S/P particles at the era of BBN with the Hubble expansion
rate of the universe. This gives
\begin{eqnarray}
\frac{5}{3} (\kappa^2\alpha_{em})\left(\frac{M_{\tilde{\gamma}}}{m_{\tilde{g}}}
\right)^2 T^3\simeq g^{1/2}_{*}\frac{T^2}{M_{P\ell}}
\end{eqnarray}
 In order for nucleosynthesis results to be uneffected (or for us to  
satisfy the bound on extra effective number of neutrinos $\Delta N_{\nu}\leq
1$ \cite{schramm}), we must have the $S/P$ as well as the gravitinos decouple
before the temperature of the universe reaches $T\simeq 200 $ MeV. Using this
value in Eq. (7), we readily deduce that $m_{\tilde{g}}\geq 1$ eV.
This is a much stronger bound on the gravitino mass than was derived in
\cite{gherghetta}.

In conclusion, for a large class of supergravity models where the 
superlight gravitino is accompanied by a superlight scalar and pseudo-scalar
particle, all gravitino masses below 50 eV are ruled out from
considerations of energy loss from the stars for $m_{S/P}\leq keV$
and those below $1~eV$ are ruled out by the BBN argument for 
$m_{S/P}\leq 1~MeV$. These results
have important implication for the scale of supersymmetry breaking because
of the intimate connection between the gravitino mass and the 
$\Lambda_{{\rm SUSY}}$ stated in the beginning. The precise lower limit
on $\Lambda_{{\rm SUSY}}$ however depends on the power $p$ in the formula,
which depends on specific supergravity model; for instance if $p\geq 2$,
we find that $\Lambda_{{\rm SUSY}}\geq 300$ TeV from the stellar bound
and $\geq 50$ TeV from the BBN bound.

\bigskip
\noindent{\Large \bf{Acknowledgements}}
\bigskip

Work of J. A. G. is partially supported by the CICYT 
Research Project AEN95-0882 and 
by the Theoretical Astroparticle Network under the EEC Contract No. 
CHRX-CT93-0120 (Direction Generale 12 COMA). The work of R. N. M.
is supported by the National Science Foundation grant no. PHY-9421386
and the work of A. R. is supported by the DOE and NASA under grant no.
NAG5-2788. One of us (A.R.) would like to thank all members of
the  High Energy Physics group at ICTP for the kind hospitality 
and friendly environment when part of this work was done. R. N. M.
would like to thank Markus Luty for discussions.


\end{document}